\documentclass{pasj00}
\draft

\begin{document}
\SetRunningHead{S. Katsuda et al.}{Suzaku Observations of SNR G156.2$+$5.7} 
\Received{2008/8/6}
\Accepted{2008/9/10}

\title{Suzaku Observations of Thermal and Non-Thermal X-Ray
  Emission from the Middle-Aged Supernova Remnant G156.2$+$5.7} 

%
 \author{%
   Satoru \textsc{Katsuda}\altaffilmark{1,2},
   Robert \textsc{Petre}\altaffilmark{1},
   Una \textsc{Hwang}\altaffilmark{1},
   Hiroya \textsc{Yamaguchi}\altaffilmark{3},
   Koji \textsc{Mori}\altaffilmark{4},
   and Hiroshi \textsc{Tsunemi}\altaffilmark{2}
}
 \email{Satoru.Katsuda@nasa.gov, Robert.Petre-1@nasa.gov,
   Una.Hwang-1@nasa.gov, hiroya@crab.riken.jp,
   mori@astro.miyazaki-u.ac.jp, tsunemi@ess.sci.osaka-u.ac.jp}
 \altaffiltext{1}{NASA Goddard Space Flight Center, Greenbelt, MD
   20771, U.S.A.} 
\altaffiltext{2}{Department of Earth and Space Science, Graduate School of
  Science, Osaka University, 1-1 Machikaneyama, Toyonaka, Osaka
  560-0043, Japan}
\altaffiltext{3}{RIKEN (The Institute of Physical and Chemical
  Research), 2-1 Hirosawa, Wako, Saitama 351-0198}
\altaffiltext{4}{Department of Applied Physics, Faculty of Engineering,
University of Miyazaki, 889-2192, Japan}


\maketitle

\begin{abstract}
We present results from X-ray analysis of a Galactic middle-aged
supernova remnant (SNR) G156.2$+$5.7 which is bright and largely
extended in X-ray wavelengths, showing a clear circular shape
(radius$\sim$50$^\prime$).  Using the Suzaku satellite, we observed
this SNR in three pointings; partially covering the northwestern rim,
the eastern rim, and the central portion of this SNR.  In the
northwestern rim and the central portion, we confirm that the 
X-ray spectra consist of soft and hard-tail emission, while in the eastern
rim we find no significant hard-tail emission.  The soft emission is
well fitted by non-equilibrium ionization (NEI) model. In the central
portion, a two-component (the interstellar medium and the metal-rich
ejecta) NEI model fits the soft emission better than a one-component
NEI model from a statistical point of view.  The relative abundances
in the ejecta component suggest that G156.2$+$5.7 is a remnant from a
core-collapse SN explosion whose progenitor mass is less than
15\,M$_\odot$.  The origin of the hard-tail emission is highly likely
non-thermal synchrotron emission from relativistic electrons.  In the
northwestern rim, the relativistic electrons seem to be accelerated by a
forward shock with a slow velocity of $\sim$500\,km\,sec$^{-1}$. 
\end{abstract}

\section{Introduction}

G156.2$+$5.7 is a large (radius$\sim$50$^\prime$), X-ray bright supernova 
remnant (SNR) discovered by the ROSAT all-sky survey (Pfeffermann et 
al.\ 1991).  The ROSAT 0.1--2.4\,keV band spectrum was well
represented by emission from a thin thermal plasma with a temperature
of about 0.5\,keV.  A Sedov analysis  
based on the X-ray data showed a very low density medium of 0.01\,cm$^{-3}$, 
an age of $\sim$26000\,yr, and a distance to the remnant of $\sim$3\,kpc.

Subsequently, successful X-ray observations were performed with Ginga 
and ASCA.  Using non-imaging instruments on board Ginga, Yamauchi 
et al.\ (1993) found that the scan profile in the 1.2--3\,keV energy 
band was similar to the soft X-ray image observed with ROSAT, while 
the scan profile above 3\,keV revealed two bumps near to the north and 
south edges of the remnant.  ASCA observed the northern and central 
portions of the remnant (Yamauchi et al.\ 1999).  ASCA spectra from 
both north and central portions were characterized by soft and hard-tail 
components.  The soft component was dominated by line emission, and 
originated from non-equilibrium ionization (NEI) plasma.  A revised 
Sedov analysis based on the spectral parameters determined by the soft 
component resulted in a low ambient density of $\sim$0.2\,cm$^{-3}$, an 
age of $\sim$15000\,yr, and a distance of $\sim$1.3\,kpc.  On the other 
hand, the hard-tail component was most likely represented by a power-law 
type spectrum, i.e., synchrotron emission from relativistic electrons.

The distance to G156.2$+$5.7 has not yet been established. Gerardy \&
Fesen (2007) discovered considerable faint H$\alpha$ line emission
coincident with X-ray emission.  Since some of them seems to be
associated with interstellar clouds located at a distance of 0.3\,kpc,
they argued that the distance to the remnant might be closer than that
was estimated previously.  We here take 1\,kpc as a distance to the
remnant, following the most recent paper on G156.2$+$5.7 (Xu et al.\
2007). 

Radio observations show a bilateral shell structure having relatively 
weak emission at the central portion of the remnant (Reich et al.\ 1992).
Although the radio surface brightness was the lowest among all known SNRs,
a magnetic field in the shell was estimated to be 16\,$\mu$G (at a 
distance of 1\,kpc) which is just slightly higher than a typical 
interstellar magnetic field.  Therefore, Reich et al.\ (1992) concluded 
that the low surface brightness was due to the low ambient density.
Polarized radio emission shows a nearly perfect tangential magnetic field 
configuration along the shell, which is typical for evolved shell-type 
SNRs (Reich et al.\ 1992; Xu et al.\ 2007).  All the observations
agree that G156.2$+$5.7 is a middle-aged SNR whose SN explosion
occurred in a low-density ambient medium.  

We focus on both thermal and non-thermal X-ray emission in this SNR,
using the most recent X-ray astronomy satellite Suzaku (Mitsuda et
al.\ 2007).  The fact that non-thermal X-ray emission is detected from
this middle-aged SNR is very interesting, since a currently accepted
theory of diffusive shock acceleration requires high shock velocity of
at least $\sim$2000\,km\,sec$^{-1}$ which is not expected in
middle-aged SNRs to produce non-thermal emission in the X-ray domain.
Fortunately, the 
thermal emission from this SNR allows us to investigate ambient medium
densities as well as the shock velocities which are 
important to study the theory of diffusive shock acceleration.
Another interesting aspect of this remnant is a clear evidence of Si
and S rich ejecta detected in the central portion of the remnant with
ASCA (Yamauchi et al.\ 1999).  We can investigate the progenitor star
which produced this SNR, by comparing the relative metal abundances of
the ejecta with those expected in nucleosynthetic models.

\section{Observations and Data Screening}

\begin{figure}
  \begin{center}
    \FigureFile(80mm,80mm){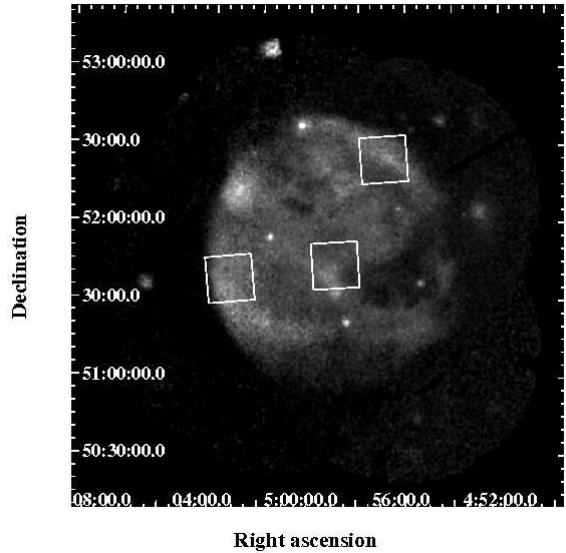}
  \end{center}
  \caption{ROSAT PSPC image of the entire G156.2$+$5.7.  The energy
    range is 0.05--3.0\,keV.  The effects of vignetting and exposure
    are corrected.  The image is binned by 15$^{\prime\prime}$ and is
    smoothed by a Gaussian kernel of $\sigma = 45^{\prime\prime}$.  The
    intensity scale is square root.  The Suzaku/XIS FOV
    (17$^\prime$.8$\times$17$^\prime$.8) are shown as white boxes.}
	\label{fig:ROSAT} 
\end{figure}

We observed the eastern (E) rim, the northwestern (NW) rim, and the 
central portion of G156.2$+$5.7 during 2007 February 16--20 with the Suzaku
satellite.  We here concentrate on the data taken
by the X-ray Imaging Spectrometer (XIS; Koyama et al.\ 2007) onboard
Suzaku.  The XIS consists of two front-illuminated (FI; XIS0 and XIS3)
CCD cameras and one back-illuminated (BI; XIS1) CCD camera.  Each
camera covers an identical imaging area of
17$^\prime$.8$\times$17$^\prime$.8.  The fields of view (FOV) of XIS 
for the three observations are shown as white boxes in the entire X-ray
image of the SNR obtained by the ROSAT PSPC (figure~\ref{fig:ROSAT}).
We reprocessed the XIS data, using the latest CTI calibration file of
version 20080131.  We cleaned the reprocessed data by the standard
criteria\footnote{See the Suzaku Data Reduction Manual which can be
  found from http://heasarc.gsfc.nasa.gov/docs/suzaku/analysis/abc.}
recommended by the calibration team of Suzaku/XIS.  After the 
screening, the remaining exposure times were 53.3\,ks for the E rim,
50.5\,ksec for the NW rim, and 51.2\,ksec for the central portion,
respectively.  

\begin{figure}
  \begin{center}
    \FigureFile(65mm,65mm){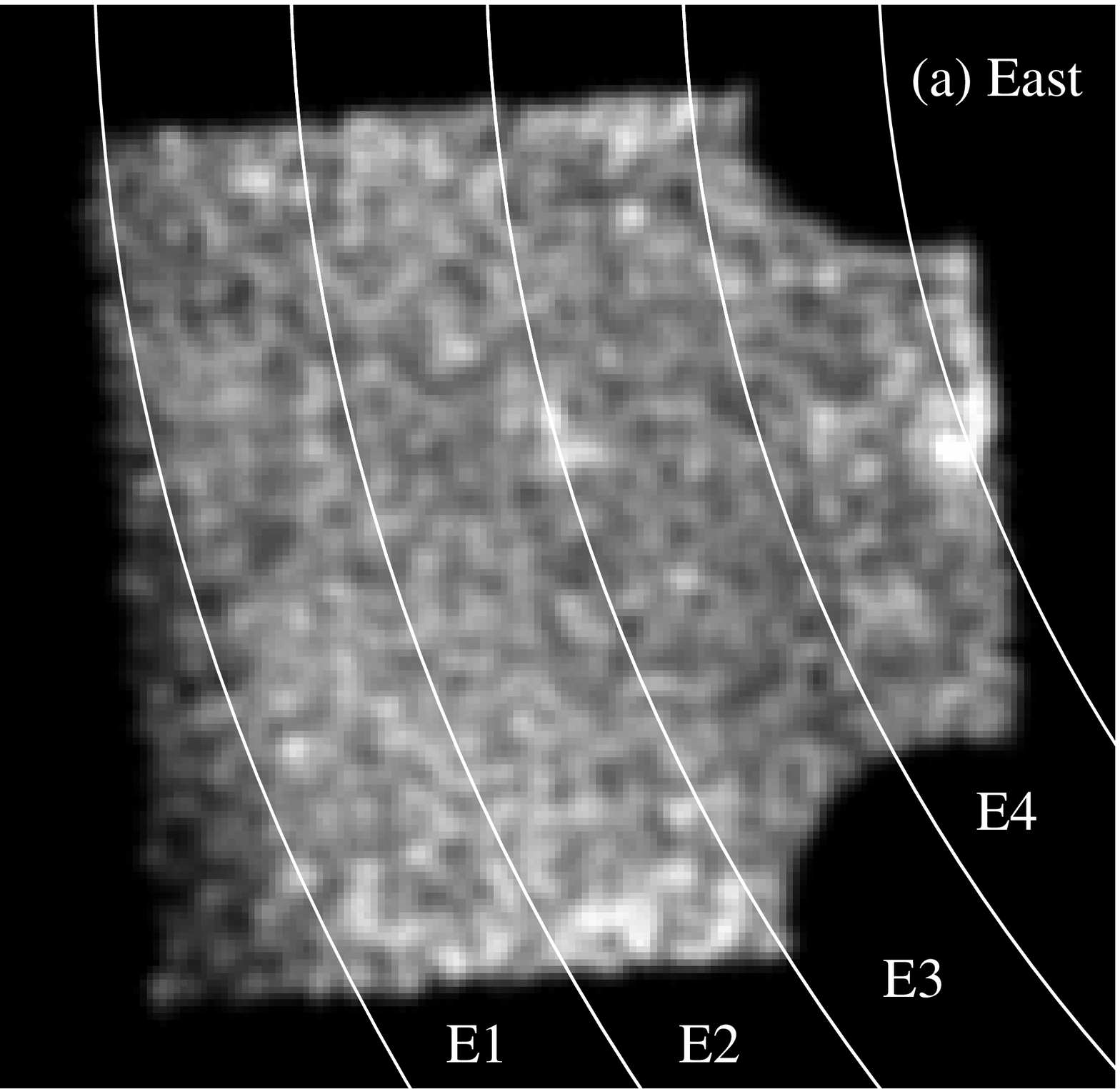}
    \FigureFile(65mm,65mm){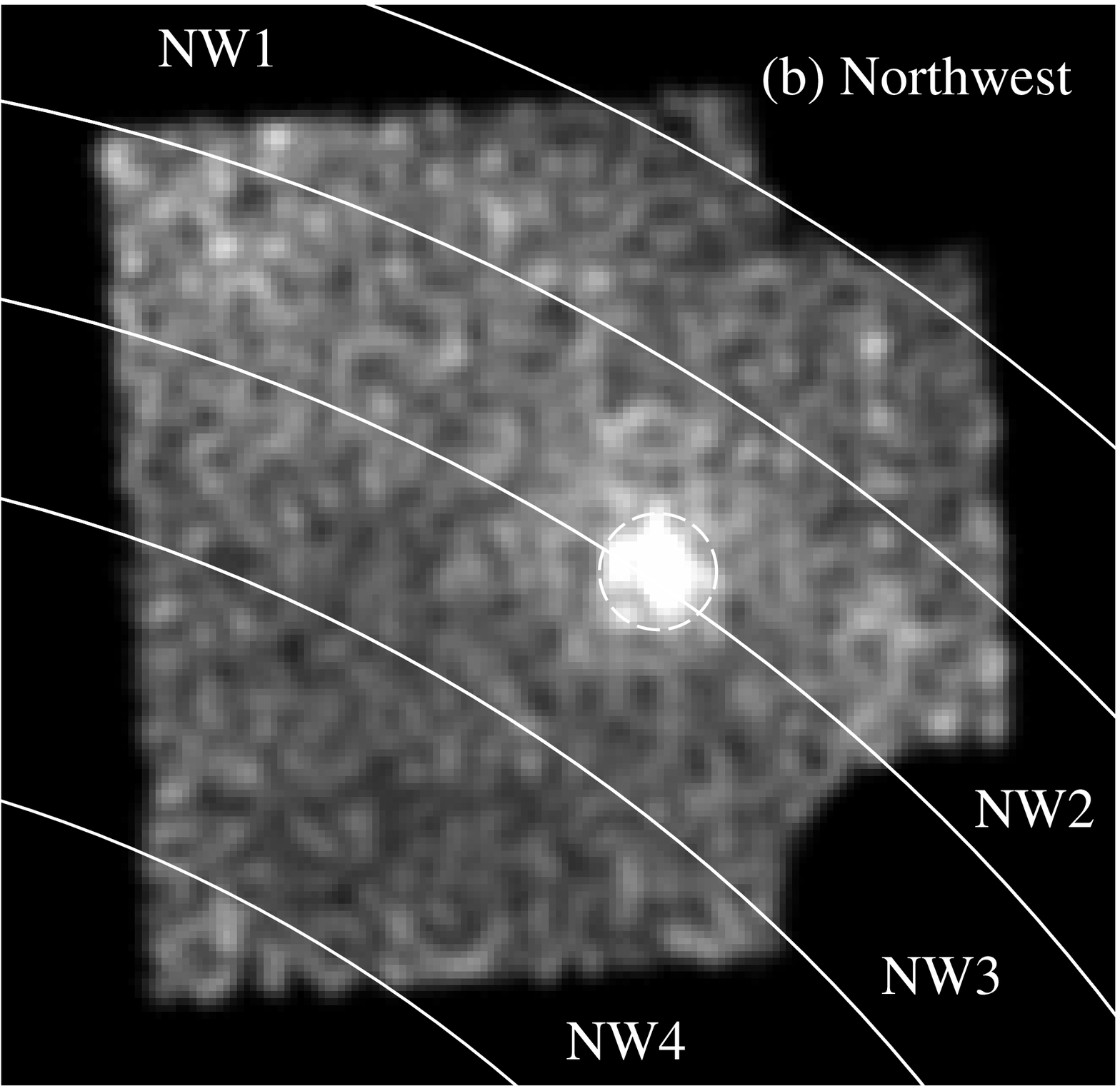}
    \FigureFile(65mm,65mm){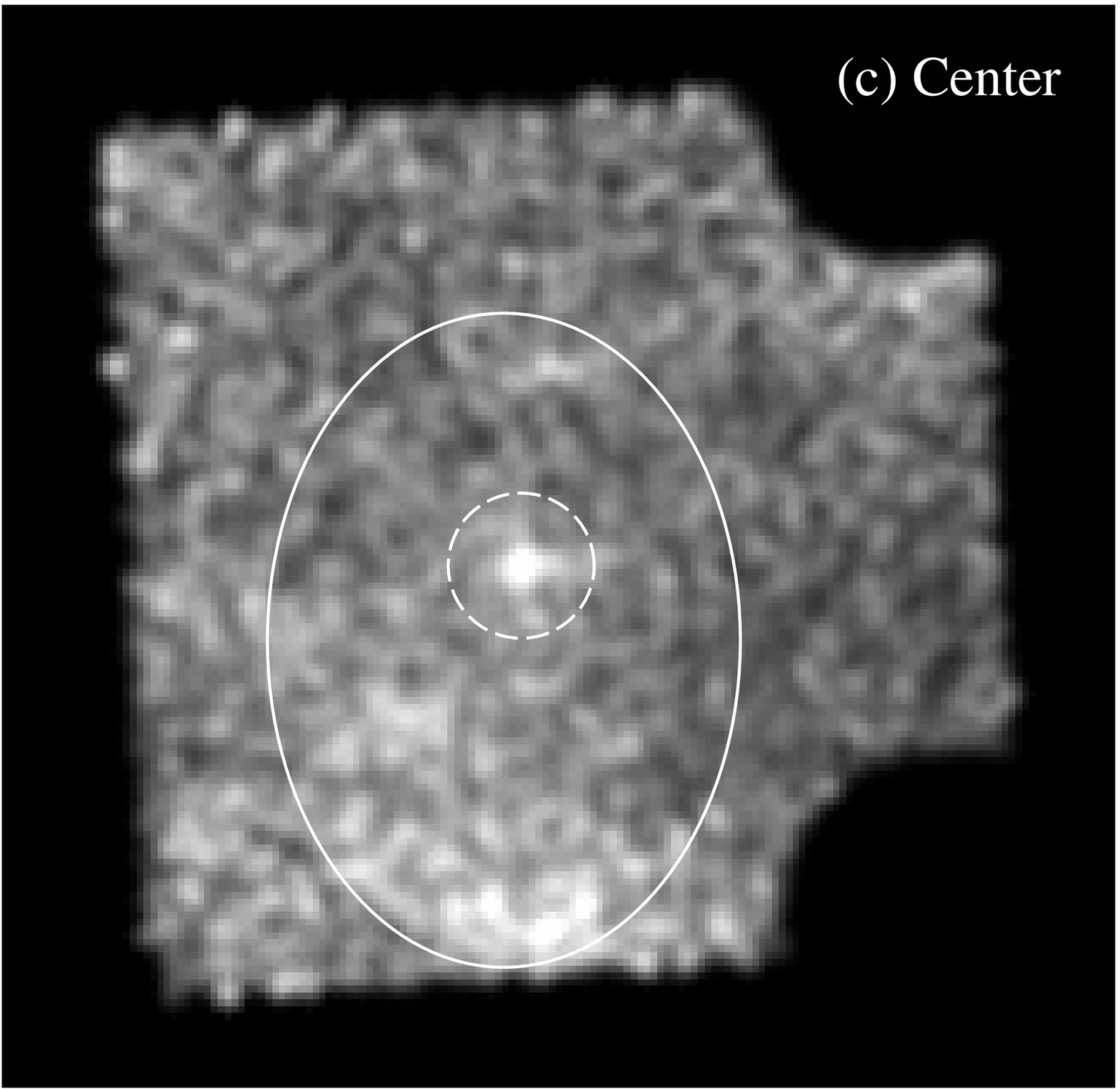}
  \end{center}
  \caption{NXB-subtracted Suzaku XIS0 images of the three pointings.
    The energy range is 0.4--8.0\,keV.  The effects of exposure,
    vignetting, and contamination are corrected. 
    The image is binned by 8$^{\prime\prime}$ and is smoothed by a
    Gaussian kernel of $\sigma = 24^{\prime\prime}$.  The intensity
    scale is square root.  White curves are the regions where we
    extract spectra.  We exclude two point sources enclosed by dashed
    circles from our spectral analyses. 
	}
	\label{fig:xis_image} 
\end{figure}

Figure~\ref{fig:xis_image} shows the vignetting-corrected XIS0 images
of the three observations.  Two corners of each FOV where the
calibration source of $^{55}$Fe is illuminated are masked in these
images.

\section{Spectral Analysis and Results}

Figure~\ref{fig:whole_spec} shows XIS0 spectra extracted from the entire
FOV for each observation.  The non X-ray background (NXB; see the next 
subsection for an explanation) spectra are also shown as red crosses.  
The signal-to-noise (black-to-red in figure~\ref{fig:whole_spec})
ratios in an energy band of 3--5\,keV significantly vary among the
three FOV, indicating a non-uniformity of the hard-tail emission
detected by Ginga and ASCA (Yamauchi et al.\ 1993; 1999).  Also
notable is a variation of equivalent widths of K-shell lines from Si
(at $\sim$1.85\,keV) and S (at $\sim$2.4\,keV); those in the central
spectrum are much larger than those in the rim spectra.  This implies
the existence of the fossil ejecta in the central portion as was
detected with ASCA (Yamauchi et al.\ 1999).  In the following spectral
fitting procedure, we use photons in the energy range of
0.4--8.0\,keV, and employ the XSPEC (version 12.4.0) software. 

\subsection{Background}

\begin{figure}
  \begin{center}
    \FigureFile(80mm,80mm){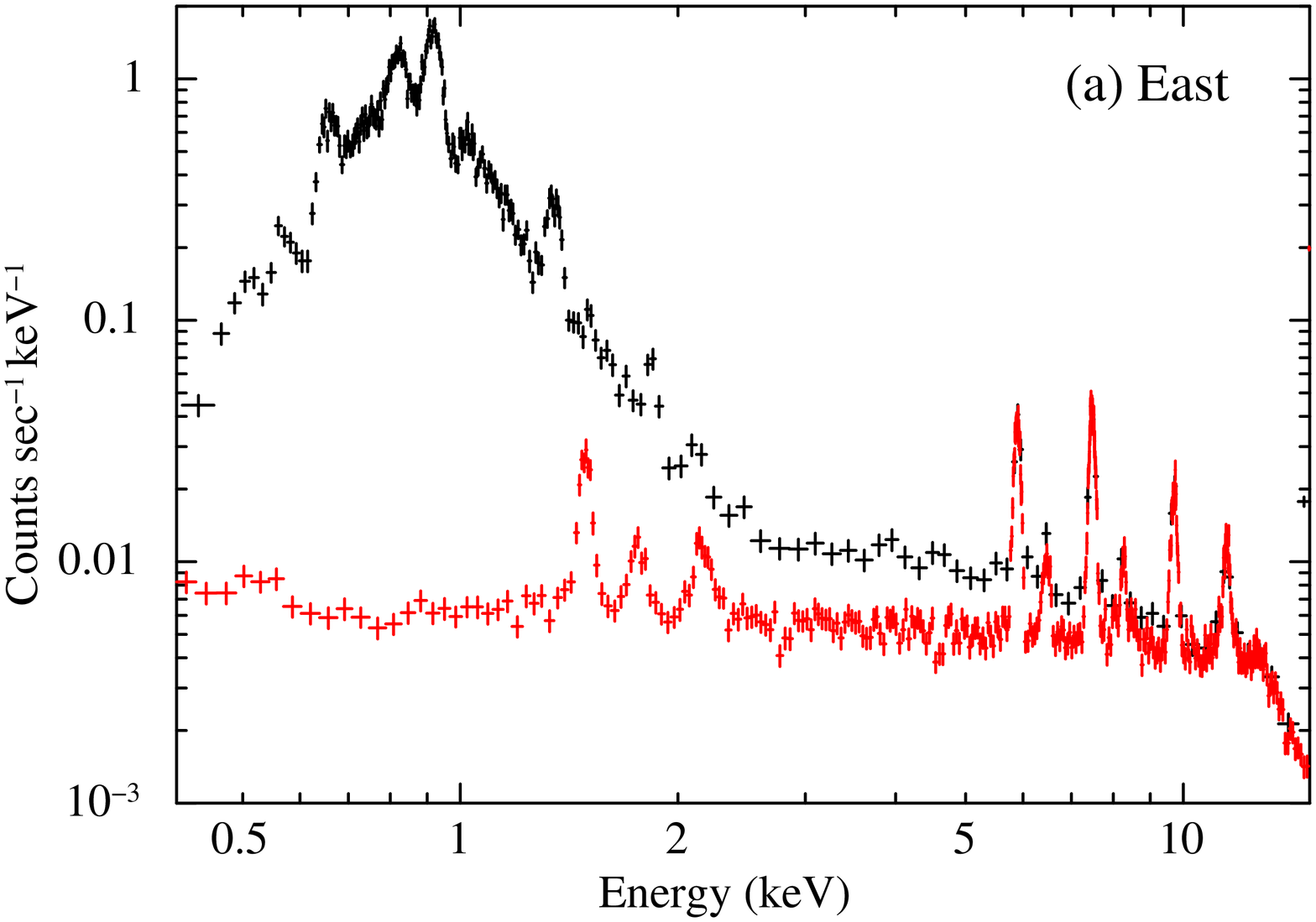}
    \FigureFile(80mm,80mm){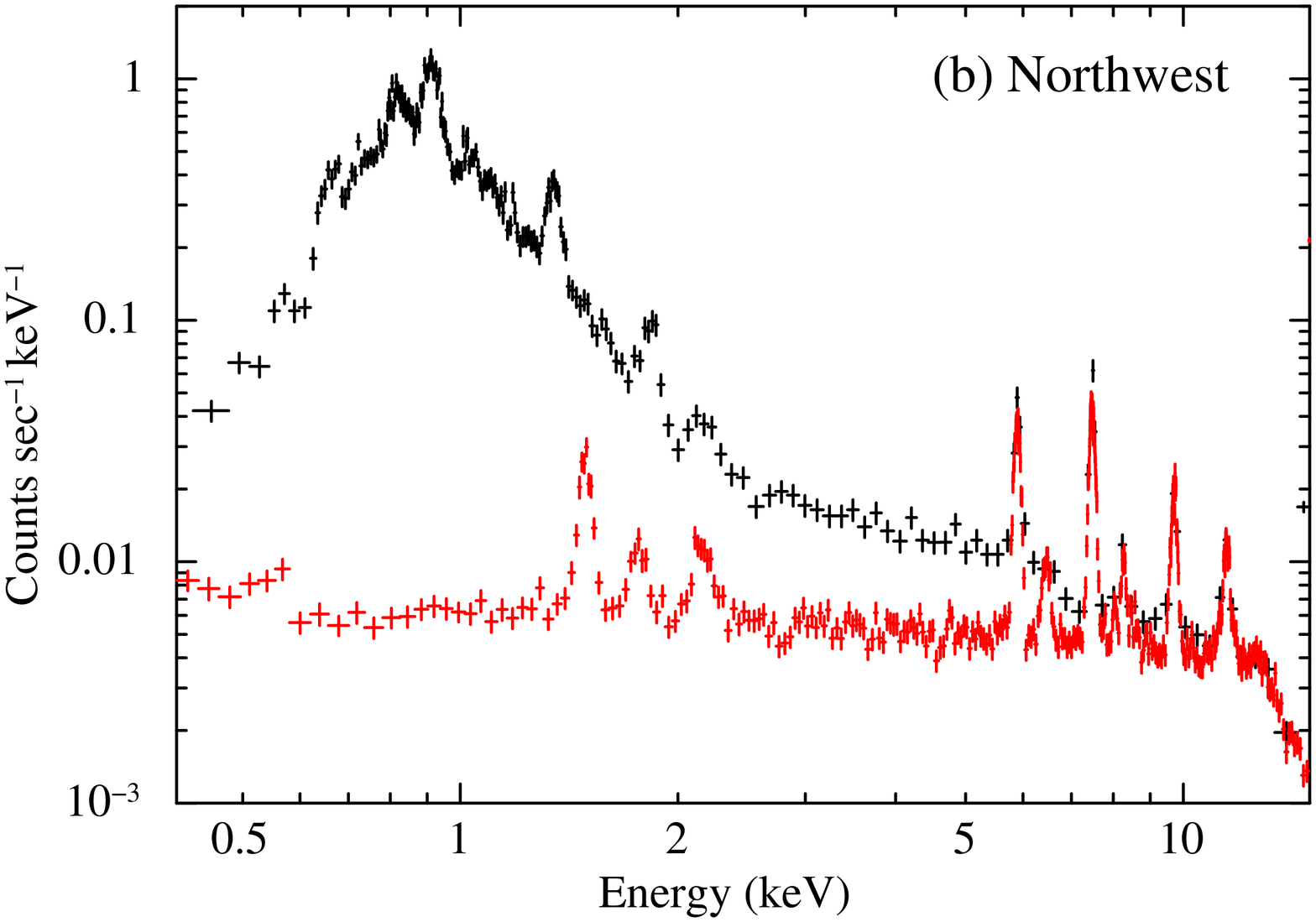}
    \FigureFile(80mm,80mm){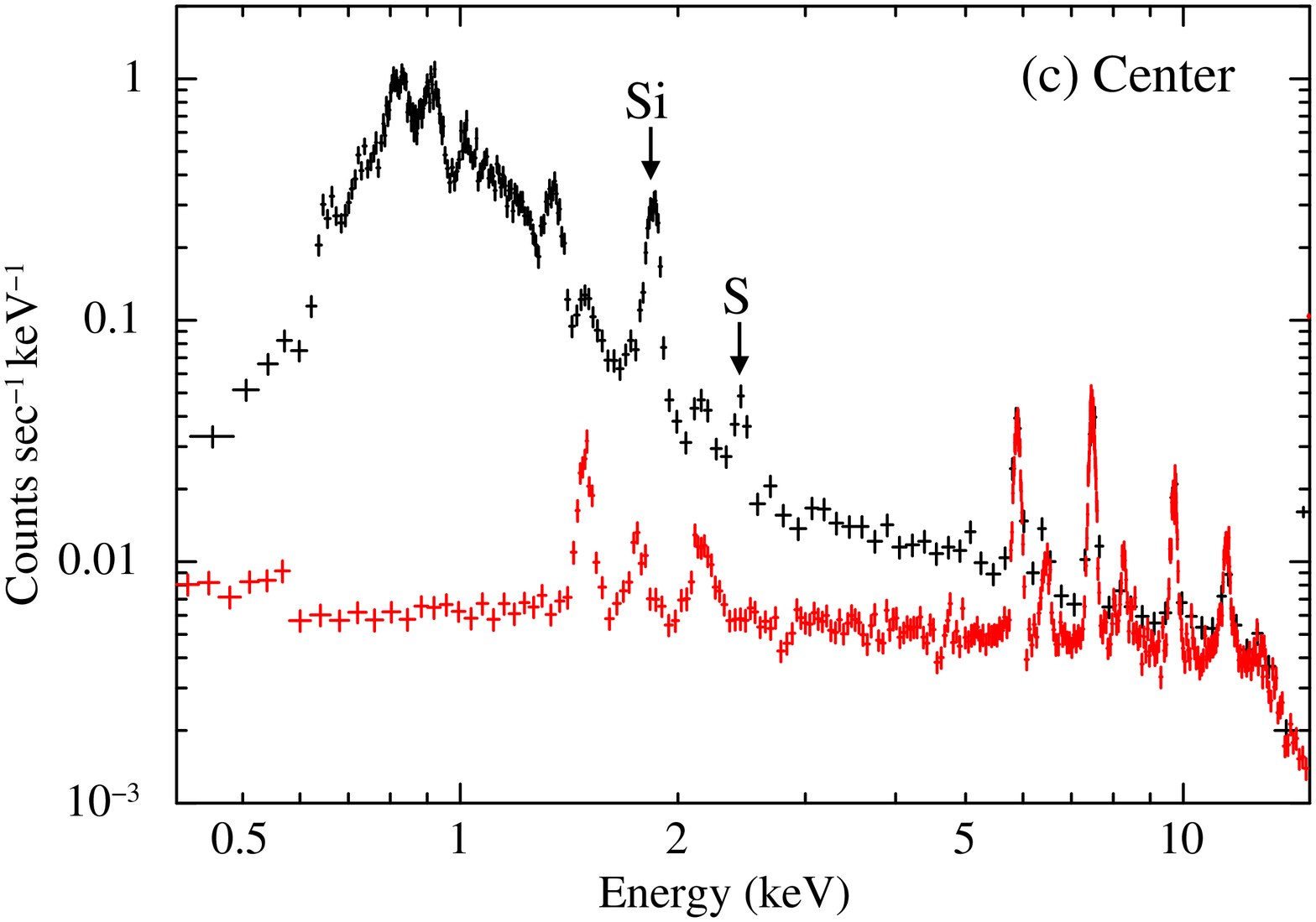}
  \end{center}
  \caption{XIS0 spectra from the entire FOV (black crosses).  
(a), (b), and (c) are responsible for the E rim, the NW rim, and the 
central portion, respectively.  Non X-ray background spectra are also 
shown as red crosses.}
\label{fig:whole_spec}   
\end{figure}

As background, we consider three kinds of sources; NXB caused by charged 
particles and $\gamma$-rays hitting the detectors, cosmic X-ray background 
(CXB) which accounts for emission from unresolved point sources, and local 
hot bubble (LHB) that is a hot gas surrounding the earth.  Note that
Galactic ridge X-ray emission (GRXE) is negligible for this SNR; based on
literature by Sugizaki et al.\ (2001) and Revnivtsev et al.\ (2006),
we estimate the flux of GRXE at the Galactic longitude, $l$, $l$ =
100$^\circ$ within Galactic latitudes, $b$, $|b| < 1^{\circ}$ to be
4$\times$10$^{-16}$\,erg\,cm$^{-2}$\,sec$^{-1}$\,arcmin$^{-2}$ in 
a range of 0.7--10\,keV, which is about 0.3\,\% of that from
G156.2$+$5.7 (Yamauchi et al.\ 1999).  Furthermore, the larger values
of $l (=156.2)$ and $b (=5.7)$ for this SNR would prefer the smaller
flux of GRXE.  We generate NXB spectra suitable for our observations,
by employing {\tt xisnxbgen} provided by the XIS team (Tawa et al.\
2008).  The generated NXB spectra are shown in
figure~\ref{fig:whole_spec}, in which we can see that the NXB spectra
above 8\,keV, where emission from astrophysical sources is negligible,
match the data.  Therefore, we are confident that NXB  
emission is properly subtracted by the NXB spectra generated by {\tt
  xisnxbgen}.  We subtract NXB emission from the same area as the
source area in detector coordinates.  For the CXB emission, we employ
broken power-law model with photon indices $\Gamma$ ($<$0.7\,keV) $=$
2.0 and $\Gamma$ 
($>$0.7\,keV) $=$ 1.4 (e.g., Miller et al.\ 2008).  As for the LHB
component, we use the APEC model (Smith et al.\ 2001) with solar
composition (Anders \& Grevesse 1989) and an electron temperature of
0.1\,keV (e.g., Tawa 2008).  We estimate the emission measure of the
LHB component around G156.2$+$5.7 to be
2$\times10^{-3}$\,pc\,cm$^{-6}$, based on the surface brightness in
the ROSAT PSPC R1 and R2 bands of $\sim$2.8
counts\,sec$^{-1}$\,arcmin$^{-2}$.  We exclude two point-like sources
visible in our FOV from our spectral analyses; the regions excluded
are enclosed by dashed circles as shown in Fig.~\ref{fig:xis_image}.

\subsection{E Rim}

We perform annular spectral analyses for the E and NW rims.  
We choose the center of the X-ray SNR at position (RA,
DEC)$=$(\timeform{4h59m05s}, \timeform{51D50'57''}) [J2000.0]
determined by a ROSAT PSPC surface brightness contour.  For the E rim,
the annular regions are spaced by 4$^\prime$ from a radial position, r
$=48^\prime$ to r $=32^\prime$, resulting in four annular regions
(from E1 at the outermost region to E4 at the innermost region).
These annular regions are shown in figure~\ref{fig:xis_image} as white
curves.  

\begin{figure}
  \begin{center}
    \FigureFile(160mm,160mm){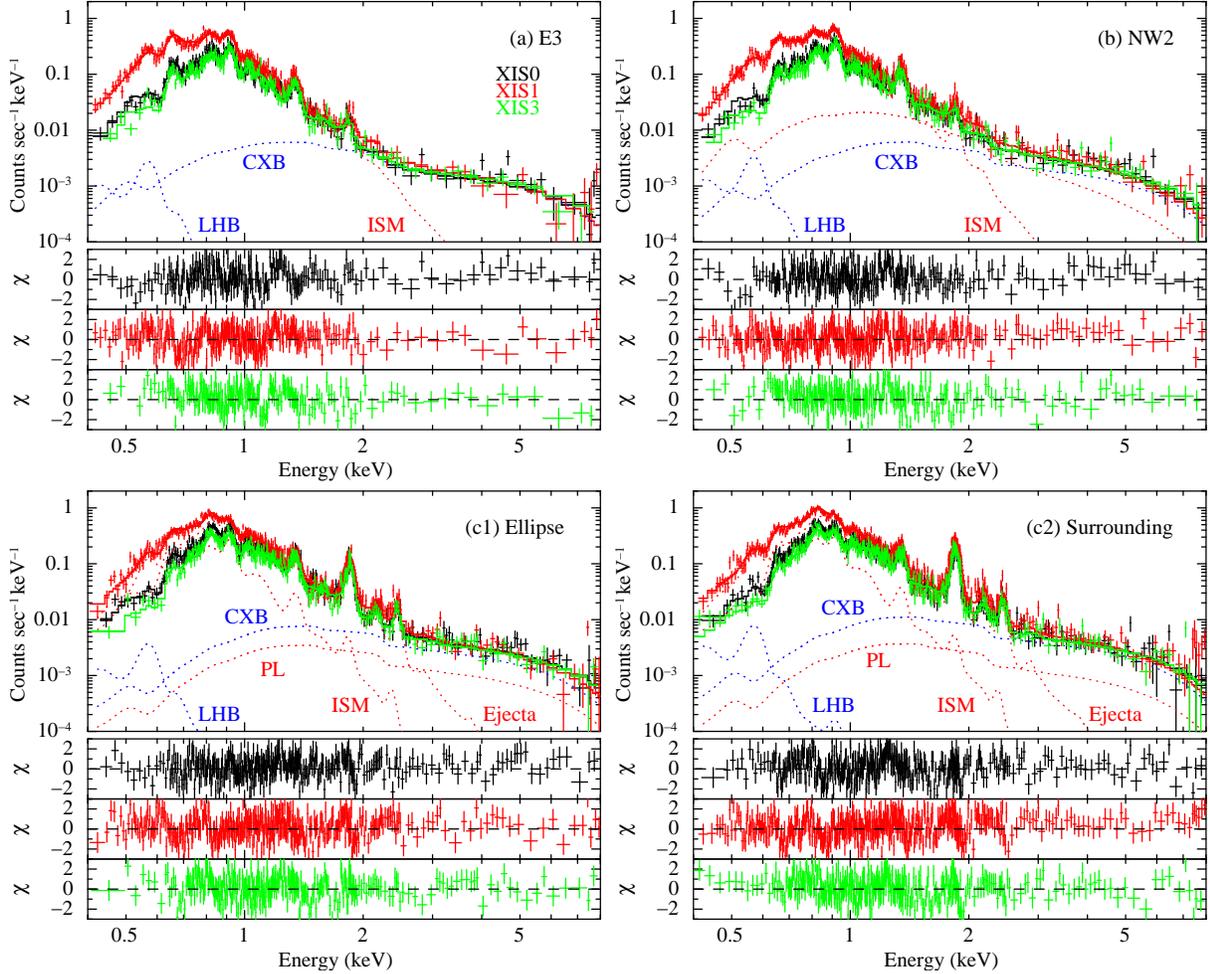}
  \end{center}
  \caption{Example spectra with their best-fit models.  The best-fit 
curves are shown as solid lines (total components) or dotted lines
(individual components).  The lower panels show the residuals.  The
spectral extraction regions are (a) from E3 in
figure~\ref{fig:xis_image} (a), (b) from NW2 in 
figure~\ref{fig:xis_image} (b), and (c1) or (c2) from the ellipse
region or the surrounding region in figure~\ref{fig:xis_image} (c).
}\label{fig:ex_spec}   
\end{figure}

We perform spectral fittings for the NXB-subtracted spectra.
Since previous ASCA observations showed non-equilibrium ionization (NEI)
states in this SNR, we apply an absorbed NEI model (VNEI model [e.g., 
Borkowski et al.\ 2001]) having LHB and CXB components for all the four 
spectra.  The free parameters are the hydrogen column density,
$N_\mathrm{H}$; electron temperature, $kT_\mathrm{e}$; the ionization
timescale, $\tau$ ($\tau$ is the product of the electron density
multiplied by the time after the shock heating); the emission measure,
EM (EM$=\int n_\mathrm{e}n_\mathrm{H} dl$, where $n_\mathrm{e}$ and
$n_\mathrm{H}$ are the number densities of electrons and protons,
respectively and $dl$ is the plasma depth); abundances of N, O, Ne,
Mg, Si, S, and Fe.  The other elemental abundances are assumed to the
solar values (Anders \& Grevesse 1989), since emission lines from
these elements cannot be seen in the Suzaku spectra. We first allow
the normalization of the CXB component to vary.  We then find that the 
mean flux of the CXB component derived in the four annular regions is
4.7$\times10^{-15}$ erg\,cm$^{-2}$\,sec$^{-1}$\,arcmin$^{-2}$ in
2.0--10.0\,keV.  This is consistent with the CXB flux of
4.4$\times$10$^{-15}$ erg\,cm$^{-2}$\,sec$^{-1}$\,arcmin$^{-2}$
estimated by blank sky observations with Suzaku (Tawa 2008).  Since
the CXB flux fluctuates by about 10\% from location to location, we
take the mean CXB flux derived in the four annular regions as a local
flux around this SNR.  After fixing the normalization of the CXB to
the mean flux, we fit all the spectra with a one-component 
VNEI model.  This model gives us fairly good fits for all the spectra;
the reduced-$\chi^2$ values (d.o.f.) range from 1.04 (753) to 1.27
(837).  No additional power-law component for possible hard-tail
emission is required at a 99\% confidence level.  Therefore, we
consider that this model well represents the spectra in the E rim.
We use this CXB flux value in the following analysis.
Figure~\ref{fig:ex_spec} (a) shows the example spectrum (from E3) with
the best-fit model.  The best-fit parameters for the four spectra are
listed in Table~\ref{tab:param1}. 

We find that the values of $kT_\mathrm{e}$ significantly decreases
toward the shock front from 0.44\,keV to 0.28\,keV.  This trend is
generally seen either in Sedov-phase SNRs or behind the shock fronts
interacting with dense 
interstellar medium (ISM) (e.g., Miyata \& Tsunemi 1999; Levenson et
al.\ 2005; Hwang et al.\ 2005).  The measured abundances are almost
constant in the FOV.  The typical abundance of 0.5 times the solar
value is consistent with those derived in the northern rim of this SNR
obtained with ASCA (Yamauchi et al.\ 1999).  The sub solar abundance
leads us to consider that the origin of the plasma in the E rim is the
ISM swept-up by the forward shock.  We should note that the N
abundance is depleted to at least 0.1 times the solar value which is
not expected in the ISM plasma.  In our data, it is quite difficult to
establish whether or not this strong depletion is real, given the fact
that we do not know the composition of contaminants built up on the
optical blocking filters of the XIS (Koyama et al.\ 2007) in addition
to weak N emission lines from G156.2$+$5.7.  The value of
$N_\mathrm{H}$ is almost constant in the FOV, and is consistent with
those derived by the ASCA observations (Yamauchi et al.\ 1999).

\begin{table*}
  \begin{center}
  \caption{Spectral-fit parameters for the E and NW rim.}\label{tab:param1} 
    \begin{tabular}{lccccc}
      \hline\hline
Parameter & E1 & E2 & E3 & E4 & NW4 \\
\hline
$N_\mathrm{H}$[$\times10^{21}$cm$^{-2}$] \dotfill &3.7$\pm$0.1& 3.2$\pm$0.1& 2.5$\pm$0.1&3.4$\pm$0.1 & 4.1$\pm$0.1\\
$kT_\mathrm{e}$[keV] \dotfill &0.28$\pm$0.01 &0.35$\pm$0.01&0.45$\pm$0.01& 0.44$\pm$0.01&0.51$\pm$0.01\\
N \dotfill& 0.11$\pm$0.05&$<$0.06&$<$0.02&$<$0.03&$<$0.04\\
O \dotfill& 0.38$\pm$0.01&0.43$\pm$0.01&0.38$\pm$0.01&0.32$\pm$0.01&0.32$\pm$0.02\\
Ne \dotfill& 0.68$\pm$0.02&0.69$\pm$0.02&0.60$\pm$0.02&0.40$\pm$0.02&0.43$\pm$0.03\\
Mg \dotfill& 0.57$\pm$0.05&0.57$\pm$0.01&0.49$\pm$0.04&0.32$\pm$0.04&0.41$\pm$0.04\\
Si \dotfill& 0.8$\pm$0.2&0.5$\pm$0.1&0.5$\pm$0.1&0.4$\pm$0.1&0.9$\pm$0.1\\
S \dotfill& $<$1&$<$0.5&$<$1&$<$0.7&1.0$\pm$0.7\\
Fe \dotfill&0.45$\pm$0.02&0.51$\pm$0.01&0.46$\pm$0.02&0.34$\pm$0.02&0.38$\pm$0.02\\
log$(\tau /\mathrm{cm}^{-3}\,\mathrm{sec})$\dotfill&10.96$\pm$0.03&10.80$\pm$0.03&10.62$\pm$0.03&10.50$\pm$0.03&10.53$\pm$0.04\\ 
EM$^\dagger$[$\times10^{18}$ cm$^{-5}$]\dotfill&3.26$\pm$0.07&1.63$\pm$0.03& 0.72$\pm$0.02&1.22$\pm$0.04&0.74$\pm$0.02\\
Surface brightness$^\S$&3.0$\times10^{-13}$&2.4$\times10^{-13}$&1.3$\times10^{-13}$&1.9$\times10^{-13}$&1.3$\times10^{-13}$\\
$\chi^2$/d.o.f. \dotfill &781/753&1065/837& 886/763&684/569&632/603\\
      \hline
\\[-8pt]
  \multicolumn{6}{@{}l@{}}{\hbox to 0pt{\parbox{140mm}{\footnotesize
     \par\noindent 
\footnotemark[$*$]Other elements are fixed to those of solar values.\\
     The values of abundances are multiples of solar value.\\  The errors
     are in the range $\Delta\,\chi^2\,<\,2.7$ on one parameter.
\par\noindent 
\footnotemark[$\dagger$]EM denotes the emission measure $\int
     n_\mathrm{e} n_\mathrm{H} dl$. 
\par\noindent 
\footnotemark[$^\S$]In unit of
erg\,cm$^{-2}$\,sec$^{-1}$\,arcmin$^{-2}$ in a range of 0.5--10\,keV. 
}\hss}}

    \end{tabular}
  \end{center}
\end{table*}

\subsection{NW Rim}

In the NW rim, we also extract four spectra from annular regions (from 
NW1 at the outermost region to NW4 at the innermost region).  The
width of the innermost region is 6$^\prime$, while those of the other
regions are 4$^\prime$.  These annular regions cover from r
$=30^\prime$ to r $=48^\prime$, and are shown in
figure~\ref{fig:xis_image} as white curves.  We first apply the same
model as we used for the spectral modeling in the E rim, i.e., a
one-component VNEI model for the emission from G156.2$+$5.7.  This 
model gives us an acceptable fit for the innermost region (NW4).
Therefore, we consider this model as the best-representative model for
the NW4 region.  The best-fit parameters are listed in
Table~\ref{tab:param1}.  On the other hand, the fit levels for the
other three spectra are not acceptable; we see excess emission from
the best-fit models above 3\,keV.  This is at least qualitatively
consistent with the fact that Yamauchi et al.\ (1999) detected
hard-tail emission above 3\,keV in the ASCA spectra.  Therefore, we
add either a thermal (NEI) or a power-law component for the hard-tail
emission.  We set metal abundances of the additional NEI component to
0.5 times the solar value which is a typical abundance of the ISM
around G156.2$+$5.7.  The $kT_\mathrm{e}$, $\tau$, and EM for the
additional NEI component, or photon index and normalization for the
additional power-law component are treated as free parameters.  Both
VNEI$+$NEI and VNEI$+$power-law model dramatically improve the fits
for all the three spectra; e.g., the value of $\chi^2$/d.o.f. for the
NW2 region is reduced from 1240/911 to 1065/909.  The example spectrum
with the best-fit VNEI$+$power-law model for the NW3 region is shown
in figure~\ref{fig:ex_spec} (b).  The best-fit parameters for the
three spectra are shown in Table~\ref{tab:param2}.   

The surface brightness of the hard-tail component,
$\sim$5$\times10^{-15}$ erg\,cm$^{-2}$\,sec$^{-1}$\,arcmin$^{-2}$, is
consistent with that detected with ASCA/GIS in the northern portion of
the remnant (Yamauchi et al.\ 1999).  We find that the
VNEI$+$power-law model fits the data slightly better than the
VNEI$+$NEI model.  However, it is not very safe to exclude the
possibility of the VNEI$+$NEI model (i.e., the thermal origin for the
hard-tail emission) from a statistical point of view.  Thus, we consider
about astrophysical aspects. If thermal origin is the case, the high
$kT_\mathrm{e}$ of above 3\,keV suggests that G156.2$+$5.7 is as young
as historical young SNRs.  As already pointed out by Gerardy \& Fesen
(2007), this is unlikely, since there have been no historical record
of the SN explosion toward G156.2$+$5.7.  Furthermore, the spatial
variation of $kT_\mathrm{e}$, i.e., the value of $kT_\mathrm{e}$ is at
least two times higher at the outermost region than those in the inner
regions (see, Table~\ref{tab:param2}), conflicts what we expect behind
shock fronts; $kT_\mathrm{e}$ generally decreases toward the shock
front due to deceleration of the shock as well as the effect of
thermal non-equilibration between ions and electrons.  On the other
hand, if non-thermal origin is the case, the spectral flattening
towards the outer region (see, Table~\ref{tab:param2}) is consistent
with that seen at the rim of Tycho's SNR
(Cassam-Cheina$\mathrm{\ddot{i}}$ et al.\ 2007).  Therefore, we 
believe that the origin of the hard-tail emission is non-thermal
emission rather than thermal emission.

The abundances derived by the VNEI component responsible for the soft
emission from G156.2$+$5.7 are similar to those observed in the E rim,
suggesting that the origin of the plasma in the NW rim is also the
swept-up ISM.  The values of $kT_\mathrm{e}$ are also similar to those
in the E rim.  This fact implies similar forward shock velocities in
both the NW rim and the E rim.

\begin{table*}
  \begin{center}
  \caption{Spectral-fit parameters for the NW rim and the central 
  portion.}\label{tab:param2} 
    \begin{tabular}{lcccccc}
      \hline\hline
Component&Parameter & NW1& NW2& NW3 & Ellipse & Surrounding\\
\hline
\multicolumn{7}{@{}c@{}}{Model: VNEI$+$NEI}\\
\hline
Absorption&$N_\mathrm{H}$[$\times10^{21}$cm$^{-2}$] \dotfill
&3.7$\pm$0.1& 3.3$\pm$0.1&3.9$\pm$0.1&4.1$\pm$0.1&3.9$\pm$0.1\\ 
VNEI&$kT_\mathrm{e}$[keV] \dotfill &0.32$\pm$0.01&
0.40$\pm$0.01&0.43$\pm$0.01 &0.46$\pm$0.01&0.52$\pm$0.01\\ 
&N \dotfill&$<$0.03&$<$0.01&$<$0.02&$<$0.02&$<$0.01\\
&O \dotfill&0.34$\pm$0.02&0.35$\pm$0.01&0.32$\pm$0.02&0.32$\pm$0.01&0.28$\pm$0.01\\
&Ne \dotfill&0.52$\pm$0.02&0.65$\pm$0.02&0.50$\pm$0.02&0.39$\pm$0.02&0.44$\pm$0.02\\
&Mg \dotfill&0.44$\pm$0.05&0.63$\pm$0.04&0.46$\pm$0.03&0.37$\pm$0.02&0.46$\pm$0.03\\
&Si \dotfill&0.6$\pm$0.2&0.7$\pm$0.1&0.6$\pm$0.1&2.1$\pm$0.1&3.0$\pm$0.1\\
&S \dotfill&$<$0.8&$<$0.4&$<$1.2&1.7$\pm$0.4&3.8$\pm$0.6\\
&Fe \dotfill&0.37$\pm$0.02&0.47$\pm$0.02&0.38$\pm$0.01&0.50$\pm$0.01&0.59$\pm$0.01\\
&log$(\tau /\mathrm{cm}^{-3}\,\mathrm{sec})$\dotfill&10.82$\pm$0.04&
10.70$\pm$0.03&10.64$\pm$0.04&10.56$\pm$0.02&10.71$\pm$0.02\\ 
&EM$^\dagger$[$\times10^{18}$cm$^{-5}$]\dotfill&2.03$\pm$0.06&
1.14$\pm$0.02&1.17$\pm$0.03&1.21$\pm$0.02&0.63$\pm$0.01\\ 
&Surface brightness$^\S$ \dotfill&2.3$\times10^{-13}$&1.7$\times10^{-13}$&1.7$\times10^{-13}$&2.1$\times10^{-13}$&1.1$\times10^{-13}$\\ 
NEI&$kT_\mathrm{e}$[keV] \dotfill &$>$7& 3.2$^{+0.4}_{-0.5}$&3.4$\pm$0.6&6$^{+4}_{-2}$&$>$5 \\
&Metal abundance \dotfill&0.5 (fixed)&0.5 (fixed)&0.5 (fixed)&0.5 (fixed)&0.5 (fixed)\\
&log$(\tau /\mathrm{cm}^{-3}\,\mathrm{sec})$\dotfill&$>$12& $>$12&$>$12&$>$12&$>$12\\
&EM$^\dagger$[$\times10^{18}$ cm$^{-5}$]\dotfill&0.030$\pm$0.005&
0.06$\pm$0.01&0.053$\pm$0.006&0.027$\pm$0.004&0.014$\pm$0.003\\ 
&Surface brightness$^\S$ \dotfill&3.7$\times10^{-15}$&5.2$\times10^{-15}$&4.8$\times10^{-15}$&3.1$\times10^{-15}$&1.8$\times10^{-15}$\\
\multicolumn{2}{@{}l@{}}{$\chi^2$/d.o.f.}\dotfill &676/588&
1065/909&981/796&1375/1101&1796/1332\\ 
\hline
\multicolumn{7}{@{}c@{}}{Model: VNEI$+$power-law}\\
\hline
Absorption&$N_\mathrm{H}$[$\times10^{21}$cm$^{-2}$] \dotfill
&3.9$\pm$0.1&3.4$\pm$0.1& 3.9$\pm$0.1&4.0$\pm$0.1&3.9$\pm$0.1  \\ 
VNEI&$kT_\mathrm{e}$[keV] \dotfill &0.31$\pm$0.01&
0.37$\pm$0.01&0.43$\pm$0.01&0.48$\pm$0.01&0.50$\pm$0.01 \\ 
&N \dotfill&$<$0.04& $<$0.02&$<$0.04&$<$0.03&$<$0.03\\
&O \dotfill&0.36$\pm$0.02& 0.43$\pm$0.02&0.40$\pm$0.02&0.36$\pm$0.01&0.38$\pm$0.02\\
&Ne \dotfill&0.53$\pm$0.02& 0.85$\pm$0.03&0.66$\pm$0.03&0.46$\pm$0.02&0.64$\pm$0.03\\
&Mg \dotfill&0.44$\pm$0.05& 0.80$\pm$0.05&0.59$\pm$0.05&0.44$\pm$0.03&0.64$\pm$0.04\\
&Si \dotfill&0.5$\pm$0.2& 0.8$\pm$0.2&0.8$\pm$0.1&2.4$\pm$0.1&4.2$\pm$0.2\\
&S \dotfill&$<$1& $<$0.7&$<$2&2.1$\pm$0.5&5.9$\pm$0.8\\
&Fe \dotfill&0.37$\pm$0.02&0.60$\pm$0.02&0.49$\pm$0.02&0.58$\pm$0.01&0.80$\pm$0.02\\
&log$(\tau /\mathrm{cm}^{-3}\,\mathrm{sec})$\dotfill&10.84$\pm$0.04&
10.79$\pm$0.03&10.67$\pm$0.04&10.58$\pm$0.02&10.74$\pm$0.02\\ 
&EM$^\dagger$[$\times10^{18}$ cm$^{-5}$]\dotfill&2.25$\pm$0.06&
1.05$\pm$0.02&0.90$\pm$0.02&1.00$\pm$0.02&0.49$\pm$0.01\\ 
&Surface brightness$^\S$ \dotfill&2.3$\times10^{-13}$&1.8$\times10^{-13}$&1.6$\times10^{-13}$&2.0$\times10^{-13}$&1.1$\times10^{-13}$\\
Power-law&Photon Index\dotfill&1.7$^{+0.2}_{-0.1}$&
2.9$^{+0.1}_{-0.9}$&2.6$\pm$0.1&2.3$\pm$0.1&2.6$\pm$0.2\\ 
&Norm$^\ddagger$ \dotfill&7$\pm$1& 38$\pm$3&27$\pm$2&13$\pm$2&13$\pm$2\\
&Surface brightness$^\S$&3.9$\times10^{-15}$&9.9$\times10^{-15}$&7.9$\times10^{-15}$&4.5$\times10^{-15}$&3.5$\times10^{-15}$\\
\multicolumn{2}{@{}l@{}}{$\chi^2$/d.o.f.}\dotfill &674/589&
1036/909&954/796&1375/1101&1781/1332\\ 
\hline
\\[-8pt]
  \multicolumn{7}{@{}l@{}}{\hbox to 0pt{\parbox{140mm}{\footnotesize
     \par\noindent 
\footnotemark[$*$]Other elements are fixed to those of solar values.\\
     The values of abundances are multiples of solar value.\\  The errors
     are in the range $\Delta\,\chi^2\,<\,2.7$ on one parameter.
\par\noindent 
\footnotemark[$\dagger$]EM denotes the emission measure $\int
     n_\mathrm{e} n_\mathrm{H} dl$. 
\par\noindent 
\footnotemark[$\ddagger$]In unit of
photons\,cm$^{-2}$\,sec$^{-1}$\,keV$^{-1}$\,sr$^{-1}$ at 1\,keV. 
\par\noindent 
\footnotemark[$^\S$]In unit of
erg\,cm$^{-2}$\,sec$^{-1}$\,arcmin$^{-2}$ in a range of 0.5--10\,keV. 
}\hss}}

    \end{tabular}
  \end{center}
\end{table*}

\subsection{Center}

We extract two spectra in the central FOV.  One is a central ellipse 
region, where we can see enhanced surface brightness relative to the
surrounding region in the ROSAT PSPC image (see,
figure~\ref{fig:ROSAT}), the other is the rest of the FOV.  
We first apply the same one-component VNEI model as we did for the
E rim spectra.  Then, we obtain the reduced-$\chi^2$ values (d.o.f.)
for the central ellipse and the surrounding region to be 1.33 (1103)
and 1.38 (1334), respectively.  Taking into consideration that
systematic uncertainties of the response, the fit level might be
acceptable from statistical point of view.  However, apparent excess
emission in the hard energy band (above 3\,keV) is found.  Therefore,
we add either an NEI or a power-law component for the hard-tail
emission.  The best-fit parameters are shown in
Table~\ref{tab:param2}. Although the additional component (either NEI
or power-law) significantly improves the fits for both spectra
(reduced-$\chi^2$ values are 1.25 for the ellipse region and 1.34 for
the surrounding region), we still see obvious discrepancy between our
data and the best-fit models around 0.5--0.6\,keV (He-like O K$\alpha$
line) and 1.2--1.3\,keV (Fe L).  This means that the VNEI$+$(NEI or
power-law) model is too simple to reproduce the soft (thermal)
emission from G156.2$+$5.7. 

Since Yamauchi et al.\ (1999) found metal-rich ejecta at the central
portion of the SNR, it is reasonable to consider that at least two
kinds of plasmas (the swept-up ISM and the metal-rich ejecta) exist
along the line of sight at the center of the SNR.  We thus apply a
two-component (ISM and ejecta 
component) VNEI model to represent the thermal emission from
G156.2$+$5.7.  Assuming that abundances of the ISM are uniform around
G156.2$+$5.7, we set abundances of the ISM component to the typical
values derived in the E or NW rims (N$=$0.05, O$=$0.4, Ne$=$0.6,
Mg$=$0.5, Si$=$0.6, S$=$0.6, and Fe$=$0.5 $\times$ the solar 
values, and the other metals are assumed to be the solar values).
In order to reduce the number of free parameters and obtain
  meaningful results, we also set the value of $kT_\mathrm{e}$ for the ISM
component to 0.3\,keV that is typical for the outermost regions in the
E and NW rim.  The $kT_\mathrm{e}$ for the ejecta component is left as a
free parameter.  The $\tau$ and EM are left as free parameters for
both ISM and ejecta components.  We also left the metal abundances of
O, Ne, Mg, Si, S, and Fe for the ejecta component as free parameters.
Since it is quite difficult to constrain the abundance of 
N due to poor statistics below 0.5 keV for the ejecta component, we
set the abundance of N to be equal to that of O.  In this way, we
apply the 2VNEI$+$(NEI or power-law) model for the two spectra
extracted in the central FOV. This model significantly improves the
fits.  The best-fit parameters 
are listed in Table~\ref{tab:param3}.  The example spectrum from the
surrounding region is shown in figure~\ref{fig:ex_spec} (c).  Since
this model gives us nearly acceptable fits for both spectra and there
is no strong reason to add further components, we stop introducing
additional components. 

If we assume that the hard-tail component is represented by the
thermal model, the value of $kT_\mathrm{e}$ is measured to be
unreasonably high (above $\sim$6\,keV).  Therefore, we consider that
the hard-tail emission is most likely explained by the power-law type
spectrum rather than the thermal spectrum.  The
metal abundances for the ejecta component are 
indeed higher than the solar values, confirming that it originates
from the metal-rich ejecta.  The abundances in the ejecta component
are significantly different between the central ellipse and the
surrounding region; the abundances for the central ellipse are all
lower than those for the surrounding region.  The relatively low
abundances at the central ellipse might be caused by either a mixing
of the ISM into the ejecta component or technical difficulty in
separating the ISM component and the ejecta component.

\begin{table*}
  \begin{center}
  \caption{Spectral-fit parameters for the central ellipse and the
  surrounding region in the central FOV.}\label{tab:param3}
    \begin{tabular}{lccc}
      \hline\hline
Component&Parameter & Ellipse & Surrounding  \\
\hline
\multicolumn{4}{@{}c@{}}{Model: 2VNEIs$+$NEI}\\
\hline
Absorption&$N_\mathrm{H}$[$\times10^{21}$cm$^{-2}$] \dotfill
&4.3$\pm$0.1& 4.6$\pm$0.1\\ 
VNEI (ISM)&$kT_\mathrm{e}$[keV] \dotfill &0.3 (fixed)&0.3 (fixed)\\
&log$(\tau /\mathrm{cm}^{-3}\,\mathrm{sec})$\dotfill&10.46$\pm$0.05&
10.64$^{+0.06}_{-0.05}$\\ 
&EM$^\dagger$[$\times10^{18}$cm$^{-5}$]\dotfill&1.09$\pm$0.05&
0.95$\pm$0.03\\ 
&Surface brightness$^\S$ \dotfill&1.3$\times10^{-13}$&1.1$\times10^{-13}$\\
VNEI (ejecta)&$kT_\mathrm{e}$[keV] \dotfill &0.48$\pm$0.01&
0.54$\pm$0.01 \\
&O(=N) \dotfill&0.48$\pm$0.04&2.8$\pm$0.3\\
&Ne \dotfill&0.25$\pm$0.03&0.9$\pm$0.1\\
&Mg \dotfill&0.40$\pm$0.03&1.5$\pm$0.1\\
&Si \dotfill&1.8$\pm$0.1&9.5$\pm$0.4\\
&S \dotfill&3.1$\pm$0.6&10$\pm$1\\
&Fe \dotfill&0.66$\pm$0.02&2.6$\pm$0.07\\
&log$(\tau /\mathrm{cm}^{-3}\,\mathrm{sec})$\dotfill&11.08$\pm$0.03&
11.30$\pm$0.02\\ 
&EM$^\dagger$[$\times10^{18}$cm$^{-5}$]\dotfill&0.76$\pm$0.02&
0.149$\pm$0.003\\ 
&Surface brightness$^\S$ \dotfill&1.2$\times10^{-13}$&0.74$\times10^{-13}$\\

NEI&$kT_\mathrm{e}$[keV] \dotfill &$>$7& $>$6 \\
&Metal abundance \dotfill&0.5 (fixed)&0.5 (fixed)\\
&log$(\tau /\mathrm{cm}^{-3}\,\mathrm{sec})$\dotfill&$>$12&$>$12\\
&EM$^\dagger$[$\times10^{18}$
  cm$^{-5}$]\dotfill&0.021$\pm$0.004&0.014$\pm$0.003\\  
&Surface brightness$^\S$ \dotfill&2.7$\times10^{-15}$&1.8$\times10^{-15}$\\
\multicolumn{2}{@{}l@{}}{$\chi^2$/d.o.f.}\dotfill &1281/1100&
1596/1331\\ 
\hline
\multicolumn{4}{@{}c@{}}{Model: 2VNEIs$+$power-law}\\
\hline
Absorption&$N_\mathrm{H}$[$\times10^{21}$cm$^{-2}$] \dotfill
&4.2$\pm$0.1&4.2$\pm$0.1  \\ 
VNEI (ISM)&$kT_\mathrm{e}$[keV] \dotfill &0.3 (fixed)&0.3 (fixed)\\
&log$(\tau /\mathrm{cm}^{-3}\,\mathrm{sec})$\dotfill&10.52$^{+0.06}_{-0.05}$&10.65$^{+0.07}_{-0.06}$\\ 
&EM$^\dagger$[$\times10^{18}$cm$^{-5}$]\dotfill&0.93$\pm$0.04&
0.71$\pm$0.03\\ 
&Surface brightness$^\S$ \dotfill&1.1$\times10^{-13}$&0.9$\times10^{-13}$\\
VNEI (ejecta)&$kT_\mathrm{e}$[keV] \dotfill &0.50$\pm$0.01&
0.55$\pm$0.01 \\ 
&O(=N) \dotfill&0.52$\pm$0.04&2.6$\pm$0.2\\
&Ne \dotfill&0.27$\pm$0.03&0.9$\pm$0.1\\
&Mg \dotfill&0.43$\pm$0.04&1.5$\pm$0.1\\
&Si \dotfill&1.8$\pm$0.1&8.8$\pm$0.3\\
&S \dotfill&2.9$\pm$0.5&9.3$\pm$1.2\\
&Fe \dotfill&0.68$\pm$0.02&2.36$\pm$0.07\\
&log$(\tau /\mathrm{cm}^{-3}\,\mathrm{sec})$\dotfill&11.12$\pm$0.03&
11.32$\pm$0.02\\ 
&EM$^\dagger$[$\times10^{18}$cm$^{-5}$]\dotfill&0.65$\pm$0.01&
0.15$\pm$0.003\\ 
&Surface brightness$^\S$ \dotfill&1.0$\times10^{-13}$&0.69$\times10^{-13}$\\

Power-law&Photon Index\dotfill&1.4$\pm$0.1&1.5$\pm$0.2\\ 
&Norm$^\ddagger$ \dotfill&3.9$\pm$0.7& 2.9$\pm$0.6\\
&Surface brightness$^\S$ \dotfill&2.8$\times10^{-15}$&1.9$\times10^{-15}$\\
\multicolumn{2}{@{}l@{}}{$\chi^2$/d.o.f.}\dotfill &1281/1100&
1584/1331\\ 
\hline
\\[-8pt]
  \multicolumn{4}{@{}l@{}}{\hbox to 0pt{\parbox{140mm}{\footnotesize
     \par\noindent 
\footnotemark[$*$]Abundances in the ISM component are fixed to the
typical value derived in the E and NW rim; N$=$0.05, O$=$0.4, Ne$=$0.6,
     Mg$=$0.5, Si$=$0.6, S$=$0.6, and Fe$=$0.5.
     Other elements are fixed to those of solar values.\\
     The values of abundances are multiples of solar value.\\  The errors
     are in the range $\Delta\,\chi^2\,<\,2.7$ on one parameter.
\par\noindent 
\footnotemark[$\dagger$]EM denotes the emission measure $\int
     n_\mathrm{e} n_\mathrm{H} dl$. 
\par\noindent 
\footnotemark[$\ddagger$]In unit of
photons\,cm$^{-2}$\,sec$^{-1}$\,keV$^{-1}$\,sr$^{-1}$ at 1\,keV. 
\par\noindent 
\footnotemark[$^\S$]In unit of
erg\,cm$^{-2}$\,sec$^{-1}$\,arcmin$^{-2}$ in a range of 0.5--10\,keV. 
}\hss}}

    \end{tabular}
  \end{center}
\end{table*}

\section{Discussion}

We detected line-dominated soft emission in all the three pointings
i.e., the E rim, the NW rim, and the central portion of this SNR.  
In the E and NW rims, the spectra of the soft emission show similar
shapes, and are represented by thermal emission originating from the
swept-up ISM.  On the other hand, in the central portion, the spectra
of the soft emission show an additional component corresponding to
metal-rich ejecta.  In addition to the soft emission, we detected
hard-tail emission in the NW rim and the central portion.  The origin
of this hard-tail emission is more consistent with non-thermal
synchrotron emission rather than thermal emission.

\subsection{Thermal Emission}

\begin{figure}
  \begin{center}
    \FigureFile(80mm,80mm){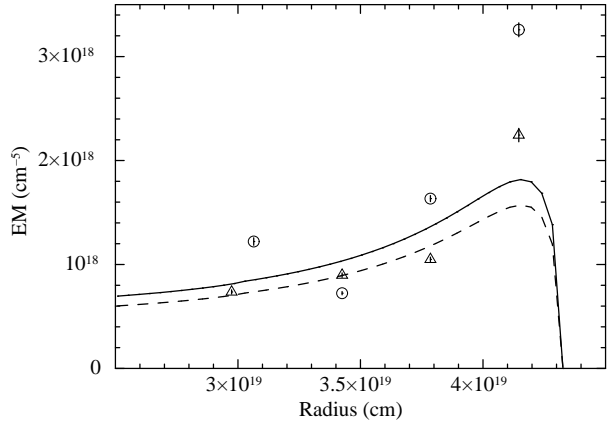}
  \end{center}
  \caption{EM profiles as a function of radius.  Our data are shown 
with circles for E rim or triangles for NW rim.  The solid or dashed 
lines are the best-representative models for E rim and NW rim, respectively.
We set the radius of the SNR to be $R=48^\prime$ (4.3$\times$10$^{19}$
cm at a distance of 1\,kpc) which is derived from the ROSAT PSPC
surface brightness contour.}
\label{fig:EM}   
\end{figure}

Our annular analyses for the rim regions revealed spatial variation of
spectral parameters.  The $kT_\mathrm{e}$ decreases from the innermost
region ($\sim$0.5\,keV) toward the shock front ($\sim$0.3\,keV), while
the EM increases toward the shock front.  These trends are at least
qualitatively consistent with those expected in the Sedov-phase SNRs.
Then, we compare the EM profiles obtained in the E and NW rims with
those derived from a simple Sedov solution, to derive physical
parameters as well as to check whether or not they are well explained
by a simple Sedov solution.  Varying the value of the ambient density
and calculating $\chi^2$-values, we search for the most suitable
(best-fit) ambient densities to represent the EM profiles for the E
rim and the NW rim, respectively.  Figure~\ref{fig:EM} shows the EM
profiles obtained in the E (data points with circles) and the NW (data
points with triangles) rims as well as the best-fit EM profiles
(solid line for the E rim and dashed line for the NW rim,
respectively) as a function of radius.  The best-fit pre-shock ambient
densities for the E and NW rims are derived to be 0.091 and 0.084
cm$^{-3}$, respectively.  These values are consistent with previous
results based on X-ray observations (0.01--0.2\,cm$^{-3}$). 
Looking at figure~\ref{fig:EM}, we see apparent deviations at the
outermost regions for both the E and NW rims; the data exceed the
best-fit models by factors of $\sim$1.8 (E) or $\sim$1.4 (NW).  This
fact suggests that the forward shock is not propagating into a uniform 
ambient density, but recently encountered a denser material
than the shock had propagated before.  The gradual increase of $\tau$
toward the shock front (see, Table~\ref{tab:param1} or
Table~\ref{tab:param2}) also supports the idea that the forward shock 
is now encountering a denser material.  

Even if the SNR is not explained by a simple Sedov solution, we can
estimate the age of the SNR by applying a simple Sedov solution in a
uniform ambient density. Then, the age is calculated to be 11000
($T_\mathrm{s}$/0.3\,keV)$^{-0.5}$($R$/50$^\prime$)($d$/1\,kpc)  
yr.  Considering that the forward shock is now encountering a denser
material, the age from the simple Sedov solution would be an upper
limit of the age of the SNR.

We found possible evidence that the forward shock is now interacting
with a denser material which surrounds the SNR.  A candidate for the
denser material is indeed found around G156.2$+$5.7 as an H{\scshape I}
shell surrounding this SNR (Reich et al.\ 1992).  Therefore, the
pre-shock ambient density of $\sim$0.09\,cm$^{-3}$ derived by the 
comparison with a simple Sedov profile must be an upper limit for the
density inside the denser material.  In this context, the SN explosion
which produced G156.2$+$5.7 occurred in a very low-density (less than
0.09\,cm$^{-3}$) medium.  The low density is reasonably expected to be
a cavity evacuated by a stellar wind of the progenitor star.  On the
other hand, the relatively dense material that the forward shock is
now encountering is suggestive to a wall of the cavity.  

\begin{figure}
  \begin{center}
    \FigureFile(80mm,80mm){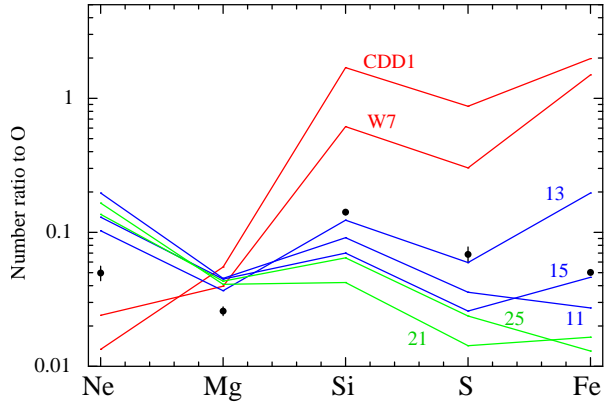}
  \end{center}
  \caption{Metal abundances of Ne, Mg, Si, S, and Fe to O.  Our data
are those derived for the ejecta component at the surrounding region
in the central FOV.  Red solid lines are those expected in 
Type-{\scshape I}a (CDD1 or W7) models (Iwamoto et al.\ 1999).  
Blue lines are those expected in core-collapse models from relatively 
low-mass progenitors (11\,M$_\odot$ [Woosley \& Weaver 1995]; 
13\,M$_\odot$ [Thielemann et al.\ 1996]; 15\,M$_\odot$ [Rauscher et al.\ 
2002]).  Green lines also represent core-collapse models but from
relatively high-mass progenitors (Rauscher et al.\ 2002).}\label{fig:abund}   
\end{figure}

A powerful method to estimate the mass of the progenitor star is
to compare relative metal abundances measured for the ejecta
with those expected in nucleosynthetic models.  We successfully
measured the metal abundances of the ejecta, eliminating the
contamination from the swept-up ISM along the line of sight by using a
two- (i.e., the ISM and the ejecta) component VNEI model.
Figure~\ref{fig:abund} shows the comparison of several metal
abundances relative to O between our data and various nucleosynthetic
models.  Here, we take the metal abundances measured in the 
surrounding region in the central FOV as the representative ejecta
abundances.  We employ Type-{\scshape I}a models from Iwamoto et al.\
(1999), while core-collapse models with progenitor masses of
11\,M$_\odot$ from Woosley \& Weaver (1995), 
13\,M$_\odot$ from Thielemann et al.\ (1996), 15\,M$_\odot$, 21\,M$_\odot$,
and 25\,M$_\odot$ from Rauscher et al.\ (2002).  We then find that our data
prefer core-collapse models with relatively low-mass progenitors (below
15\,M$_\odot$) for the origin of G156.2$+$5.7.  However, we should
note that the metal abundances of the ejecta are determined in a very
limited region of G156.2$+$5.7.  In order to determine the metal
abundances for total ejecta, we need further X-ray observations of
G156.2$+$5.7.  

The H{\scshape I} shell surrounding G156.2$+$5.7 is possible evidence
of the wall of the cavity as already suggested by Reich et al.\
(1992).  If this H{\scshape I} shell is indeed the wall of the cavity,
the radius of the cavity is estimated to be $\sim$15 ($d/1$\,kpc)\,pc
(i.e., the radius of the SNR).  McKee et al.\ (1984) calculated the
size of a cavity created by a massive star of type earlier than B0
($\sim$15\,M$_\odot$) to be a radius of 50\,pc.  Within the range of
the distance (1--3\,kpc) to G156.2$+$5.7, the radius of the cavity is
lower than 50\,pc.  Therefore, the relatively small size of the
possible cavity also supports that the progenitor star of G156.2$+$5.7
is less massive than 15\,M$_\odot$.

\subsection{Non-Thermal Emission}

In the NW rim, the non-thermal emission comes from just behind the
forward shock (see, Table~\ref{tab:param2}).  This suggests that the
non-thermal emission originates from the relativistic electrons 
accelerated by the forward shock.  Then, we should note that the value 
of $kT_\mathrm{e}$ just behind the forward shock is derived to be about
0.3\,keV, which results in a shock velocity of 
$\sim$500 ($kT_\mathrm{e}$/0.3\,keV)$^{0.5}$\,km\,sec$^{-1}$
assuming thermal equilibration between ions and
electrons.  The assumption seems to be reasonable, since the
the expected proton temperature, $\sim$0.2  
($\tau$/6$\times10^{10}$\,cm$^{-3}$sec)$^{-1}$($kT_\mathrm{e}$/0.3\,keV)$^{2.5}$\,keV,
is comparable to the measured electron temperature.  Although this
equation is valid only for highly non-equilibrium stage of shocked
plasma, the resultant value at least supports that thermal equilibrium
is almost reached.  In addition, we should note that the
electron-proton thermal equilibration is expected in slower shocks
(e.g., Rakowski 2005).  It is surprising that we detect non-thermal
X-ray emission from such a  slow shock, since a currently accepted
theory of diffusive shock acceleration requires high shock velocity of
at least 2000\,km\,sec$^{-1}$ to produce non-thermal emission in the
X-ray energy band.  The slow shock might require unacceptably high
efficiency of particle acceleration.  Therefore, we check the validity
of non-thermal origin for the hard-tail emission, based on
morphologically and spectral features.

Regarding the morphology, relatively strong non-thermal radio
emission is detected in the northern and the southern shell regions,
while relatively weak emission is detected from the central portion of
the remnant.  Therefore, the distribution of the X-ray hard-tail
emission matches the radio morphology; the X-ray hard-tail emission is 
detected with strong intensity in the NW rim, with weak intensity in
the central portion, while it is not significantly detected in the E
rim.  This suggests that the radio emission and X-ray hard-tail
emission have the same origin with each other, supporting the origin
of the X-ray hard-tail emission is non-thermal synchrotron.  In this
context, we expect to detect non-thermal X-ray emission from the
southern rim of the SNR, which will be checked by future X-ray
observations.

\begin{table*}
  \begin{center}
  \caption{Spectral-fit parameters derived by the SRCUT
  model.}\label{tab:param4} 
    \begin{tabular}{lccccc}
      \hline\hline
Parameter & NW1&NW2&NW3& Ellipse& Surrounding\\
\hline
$\nu_\mathrm{rolloff}$[$\times10^{16}$Hz]\dotfill&0.95$\pm$0.05&1.28$\pm$0.04&1.28$\pm$0.02&1.63$\pm$0.09&1.20$\pm$0.05\\  
Surface brightness$^\dagger$ \dotfill&0.75$\times10^{-14}$&1.2$\times10^{-14}$&1.2$\times10^{-14}$&0.69$\times10^{-14}$&0.43$\times10^{-14}$\\
$\chi^2$/d.o.f. \dotfill&691/590&1043/910&972/797&1315/1101&1610/1332\\
      \hline
\\[-8pt]
  \multicolumn{6}{@{}l@{}}{\hbox to 0pt{\parbox{140mm}{\footnotesize
     \par\noindent 
\footnotemark[$*$]The errors are in the range $\Delta\,\chi^2\,<\,2.7$
on one parameter. 
\par\noindent 
\footnotemark[$^\dagger$]In unit of
erg\,cm$^{-2}$\,sec$^{-1}$\,arcmin$^{-2}$ in a range of 0.5--10\,keV. 
}\hss}}

    \end{tabular}
  \end{center}
\end{table*}

To check spectral features in multi-wavelength emission, we introduce
the SRCUT model, which describes synchrotron radiation from a
power-law distribution of electrons with an exponential cut-off
(Reynolds \& Keohane 1999) instead of the power-law model.  In this model, 
we fix the radio spectral index to 0.48 (Xu et al.\ 2007).  
The flux at 1\,GHz is also fixed at the value determined by radio 
observations (based on figure~1 in Xu et al.\ 2007); 
2.1\,mJy\,arcmin$^{-2}$ for the NW rim, 0.83\,mJy\,arcmin$^{-2}$ for
the central portion.  Thus, the only free parameter in the SRCUT model
is the roll off frequency, $\nu_\mathrm{rolloff}$.  In this way, we
fit all the spectra in which we detected hard-tail emission.  The
measured roll off frequencies and $\chi^2$-values are summarized in
Table~\ref{tab:param4}.  The fits are fairly good, although they are
slightly worse than those derived by using the power-law model (see,
Table~\ref{tab:param2} or Table~\ref{tab:param3}).  The measured roll off
frequency, $\sim$1.5$\times$10$^{16}$\,Hz, is relatively low compared
with those derived in young historical SNRs;
26$\times10^{16}$Hz for Tycho (Tamagawa et al.\ 2008),
20$\times10^{16}$Hz for Kepler (Cassam-Cheina$\mathrm{\ddot{i}}$ et
al.\ 2004), and
5.7$\times10^{16}$Hz for SN1006 (Bamba et al.\ 2008).
The low value of $\nu_\mathrm{rolloff}$ is in a sense that the
efficiency of particle acceleration can be reduced.
Now, we calculate the efficiency of particle acceleration.
The roll off frequency is expressed as
\[ 
\nu_\mathrm{rolloff} =
1.6\times10^{16}(B/10\mu\mathrm{G})(E_\mathrm{max}/10 \mathrm{TeV})^2~~
\mathrm{Hz}, 
\]
where $B$ is a strength of ambient magnetic field and $E_\mathrm{max}$
is a maximum energy of accelerated electrons (Reynolds 1998).
Equating the synchrotron loss time to acceleration time, $E_\mathrm{max}$
is written down as follows (Uchiyama et al.\ 2003),
\[
E_\mathrm{max} = 67 (B/10\mu\mathrm{G})^{-0.5} 
(v_\mathrm{s}/2000\,\mathrm{km}\,\mathrm{sec^{-1}}) \eta^{-1/2}~~\mathrm{TeV},
\]
where $v_\mathrm{s}$ is the shock velocity and $\eta$ is the so-called
gyrofactor, i.e., the ratio of the particle mean free path to the
gyroradius, and represents the efficiency of particle acceleration.
Combining the above two equations, we can derive the following equation,
\[ 
\nu_\mathrm{rolloff} =
7.2\times10^{17} (v_\mathrm{s}/2000\,\mathrm{km}\,\mathrm{sec^{-1}})^2 
\eta^{-1}~~\mathrm{Hz}.  
\]
Substituting the measured $\nu_\mathrm{rolloff}$ of 1.5$\times10^{16}$\,Hz 
and the forward shock velocity of 500\,km\,sec$^{-1}$, we obtain the
value of $\eta$ to be 48 (If thermal equilibration has not achieved
yet behind the shock front, the shock velocity will be faster than
500\,km\,sec$^{-1}$ so that $\eta$ will be larger than 48).  Thus, the  
particle acceleration is much less efficient than that taken place in
the RXJ1713.7-3946 SNR, in which we see the extreme Bohm diffusion
limit, $\eta =$ 1 (Uchiyama et al.\ 2007).  Therefore, the fact that
the slow ($\sim$500\,km\,sec$^{-1}$) shock in G156.2$+$5.7 produces
non-thermal X-ray emission turns out to be acceptable from a
theoretical point of view of diffusive shock acceleration.  We
consider that the detection of non-thermal X-ray emission associated
with the slow shock in G156.2$+$5.7 is just simply because the flux
ratio of non-thermal to thermal emission in this SNR is higher than
those in the general middle-aged SNRs where non-thermal X-ray emission
is not detected.  Ellison et al.\ (2000) argued that a weak ambient
magnetic field and/or a low ambient density yield the high flux ratio
between non-thermal and thermal emission.  As for the ambient density,
we obtain similar ambient densities between the E rim
(0.093\,cm$^{-3}$) and the NW rim (0.087\,cm$^{-3}$).  However, we
did not detect significant non-thermal emission from the E rim, but
detected it only from the NW rim.  Therefore, we suggest that the
magnetic field plays an important role to the non-thermal/thermal flux
ratio for the case of G156.2$+$5.7.

We discuss the nature of the central non-thermal emission in
G156.2$+$5.7.  The central bump isolated from both the northern and
the southern shell in the radio image reminds us of a plerion for the
nature of the central non-thermal emission, although there is
no apparent counterpart of a pulsar/pulsar wind nebula.  Xu et al.\
(2007) conclude that the emission from the central portion came from
the shell of the SNR rather than a plerion because radio spectra do
not show spectral flattening in the central portion of the remnant
which is expected in the case of pulsar wind nebula. In principle, it
is possible for the shell to be projected to the center of the SNR. In
the simulations of Orlando et al.\ (2007) to investigate the origin of
the asymmetries in bilateral SNRs, we find such an example. In the
paper, we find that the radio morphology of G156.2$+$5.7 is similar to
the surface brightness of non-thermal emission expected in a situation
that the bilateral radio limbs are parallel to the ambient magnetic
field and a gradient of ambient density is perpendicular to the
magnetic field (see figure 5 in Orlando et al.\ 2007).  Although the
possibility that the central non-thermal emission is a plerion can not
be completely ruled out, it is preferred that the central non-thermal
emission is associated with the shell of the SNR. Further X-ray
observations around the central portion of the SNR will allow us to
investigate a detailed variation of spectral index around the center
of the SNR. If we see strong spectral variation, as is generally seen
in pulsar wind nebulae (e.g., Petre et al.\ 2007), the origin would be
most likely a plerion, otherwise the central non-thermal emission is
associated with the shell of the SNR.

\section{Conclusions}

We have observed the middle-aged SNR G156.2$+$5.7 with the Suzaku
satellite in three pointings which are targeted at the E rim and the
NW rim, and the 
central portion of the SNR.  Hard-tail X-ray emission is detected in
the NW rim and the central portion, while it is not significantly
detected in the E rim.  The distribution of the hard-tail emission
seems to be coincident with that of the radio non-thermal emission.
Also, multi-wavelength emission is well represented by a so-called
SRCUT model.  These facts lead us to conclude that the hard-tail
emission originates from non-thermal synchrotron emission.  The
velocity of the forward shock which likely produces the non-thermal
X-ray emission in the NW rim is estimated to be
$\sim$500\,km\,sec$^{-1}$, assuming thermal equilibrium between ions
and electrons.  The shock might be the slowest among those having
non-thermal emission in the X-ray wavelength.  The relative
abundances in the ejecta component suggests that the origin of
G156.2$+$5.7 is a core-collapse SN explosion whose progenitor is a
relatively low-mass star of $\sim$15\,M$_\odot$.

\bigskip

This work is partly supported by a Grant-in-Aid for Scientific
Research by the Ministry of Education, Culture, Sports, Science and
Technology (16002004).  This study is also carried out as part of
the 21st Century COE Program, \rq{Towards a new basic science:
depth and synthesis}\rq.  S.K.\ is supported by a JSPS Research
Fellowship for Young Scientists.  S.K.\ is also supported in part by
the NASA grant under the contract NNG06EO90A.



\end{document}